# A Novel Piezoelectric Microtransformer for Autonmous Sensors Applications

Patrick SANGOUARD, Gaëlle LISSORGUES and Tarik BOUROUINA, *Senior Member, IEEE*

Université Paris-Est, ESIEE, Laboratoire ESYCOM, EA 2552
2 Bd Blaise Pascal 93162, Noisy-le-Grand, France

*Abstract-* **This work relates to a novel piezoelectric transformer to be used in an autonomous sensor unit, possibly in conjunction with a RF-MEMS retro-modulator.**

## I. INTRODUCTION

This works relates to autonomous devices like sensors. The concept refers to sensors or actuators without any embedded power source. These are positioned for instance inside a wall, a human body or in space. Such devices are passive most of the time, but must provide, on request, information to the outside world or perform actions. Wireless and with no internal energy source, the energy needed for the operation of these autonomous devices must therefore be provided by the extern world. One can perform a transformation of this external energy, which can be originally ultrasonic or in the form of electromagnetic waves, to an electrical energy.

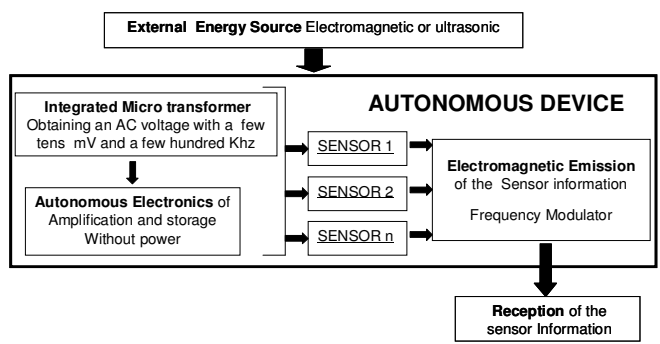

Fig. 1. Illustration of the proposed concept of autonomous device

## II. OPERATION PRINCIPLE

The micro transformers have to transform an electromagnetic (or an ultrasonic) energy into an electrical energy. Their operation is based on some properties of piezoelectric thin films. They also must comply with the constraints associated with the ultimate goal of integration on the same silicon wafer of the sensors (or actuators), the frequency modulator and the electronics circuit. The shape of the micro-transformers is necessarily planar as they are created by micro-machining of thin- piezoelectric film of a few microns in thickness. The exploitation of thickness distortion is very difficult because in this case the resonance frequency of the piezoelectric structures quickly becomes excessive (several tens of Megahertz). On the other hand, the planar structures, the corresponding deposition processes and the layers polarization do not allow directly to use the transverse and shearing piezoelectric effects ($d_{13}$ and $d_{15}$.piezolectric coefficients, respectively)

Taking into account these constraints, we are developing micro transformers using some resonance modes of bending deformation. These are better suited to micro systems. The main interest of this solution is that we can control the resonance frequency by modifying the geometrical dimensions of the structure.

## III. FIRST GENERATION OF PIEZOELECTRIC TRANSFORMERS

In a previous work [1-3] we made a first generation of micro transformers, which take advantage of the direct and reverse piezoelectric effect and a mechanical amplification, taking place between the center and the boundary of a vibrating electromechanical structure (Fig. 2).

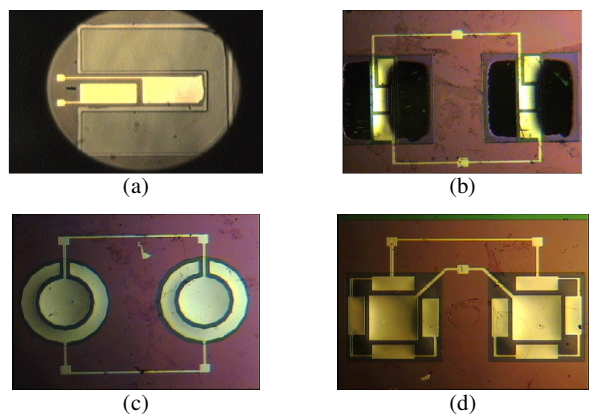

(a)     (b)
(c)     (d)

Fig. 2. Photographs of fabricated micro transformers. Direct and reverse piezoelectric effects take place at the center and at the boundaries of mechanical structures, where sets of electrodes are deposited. The different strain level in these regions is used to obtain the required voltage amplification in the transformers

These first versions of micro transformers have been done with simple geometric configurations and took the form of beams, bridges and circular or square discs.

A modelling of these devices has been established [4] by splitting these structures into simple elements, one formed by the primary (input) of the micro transformer, the other by his secondary (output). These elements are connected by boundary conditions involving efforts (equal bending moments, and cutting efforts), as well as speed (equal linear and angular velocity). Calculation details are given in the Appendix.

The equivalent diagram (Fig. 4) is deduced from this analytical model of the beam micro transformer, which consists here of a single film piezoelectric, which is common to the primary and the secondary electrodes.

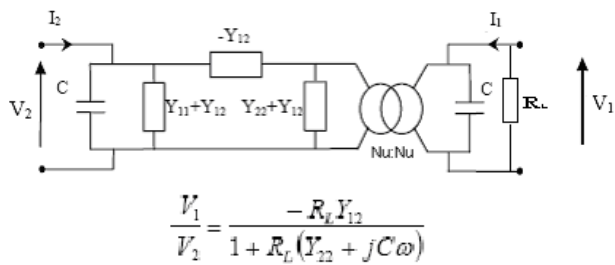

Fig. 4: Equivalent electric circuit and gain a beam micro transformer consists of one piezoelectric film

The previous model was validated on a macro transformer using a single film piezoelectric (shown in Fig.5), whose geometrical details and physical properties are given in Table 1 below.

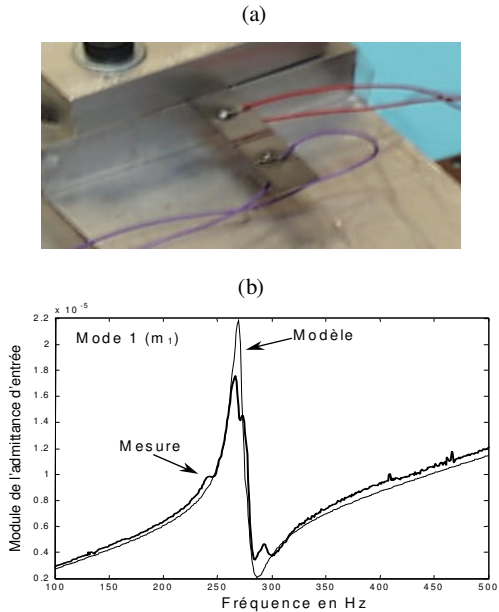

Fig. 5: Macroscopic piezoelectric transformer (15mm* 6.3 mm)
(a) Picture (b) Comparison between measurements and theoretical model

| L | w | $h_p$ | $h_m$ | $\rho_p$ |
|---|---|---|---|---|
| 15mm | 6,35mm | 0,1905mm | 0,0635mm | 7800kg/m³ |
| $d_{31}$ | $s_{p11}$ | $s_{m11}$ | $\varepsilon^T_{33}$ | $\rho_m$ |
| $-190 \cdot 10^{-12}$ | $15,15 \cdot 10^{-12}$ | $5 \cdot 10^{-12}$ | $1800\varepsilon_0$ | 2690kg/m³ |

Table 1: Characteristics of the macroscopic piezoelectric transformer

The characteristic of the first micro transformers of simple geometric shape, made on a silicon wafer with a single piezoelectric film of either PZT or AlN (Fig. 6) were consistent with the theory developed in thesis [4], but the gain measured electrical thereof were quite weak.

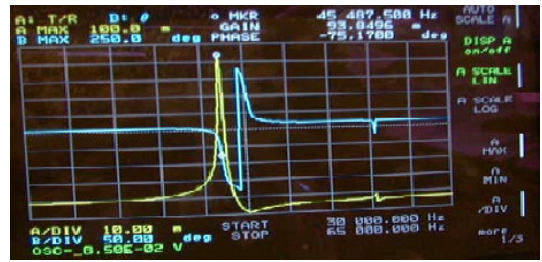

Fig. 6: Gain (0.093) and phase of a beam micro transformer using an uniform AlN film on the primary (input) and secondary (output) electrodes

IV. SECOND GENERATION OF MICROTRANSFORMERS WITH TWO PIEZOELECTRIC MATERIALS

Taking into account these first results, and in order to improve the voltage gain, we explored a new idea, which consists of using two different piezoelectric films, called "complementary". Indeed, the first piezoelectric material is chosen with a high piezoelectric coefficient - like PZT (Lead Zirconate Titanium), whose $d_{31}$ is typically equal to $1.8 \cdot 10^{-10}$ m/V. This first material is used in the primary (input) electrode). The second piezoelectric coefficient is chosen for that it has a much lower piezoelectric coefficient - such as AlN (Aluminium Nitride), whose $d_{31}$ is typically equal to $2.65 \cdot 10^{-12}$ m/V. This second material is used on the secondary (output) electrode

The comparative gains simulations with Matlab software for identical micro transformers structures are shown in Fig. 7, for micro-transformers based on a bridge-type resonator and in Fig. 8 for micro-transformers based on disc-type resonators, respectively. These two kinds of micro-transformers are compared in two configurations, where we used either the same piezoelectric film (AlN) on the primary and secondary electrodes (Figs. 7.a and 8.a) or two "complementary" piezoelectric films : PZT and AlN (Figs. 7.b and 8.b) on the primary and secondary electrodes, respectively.

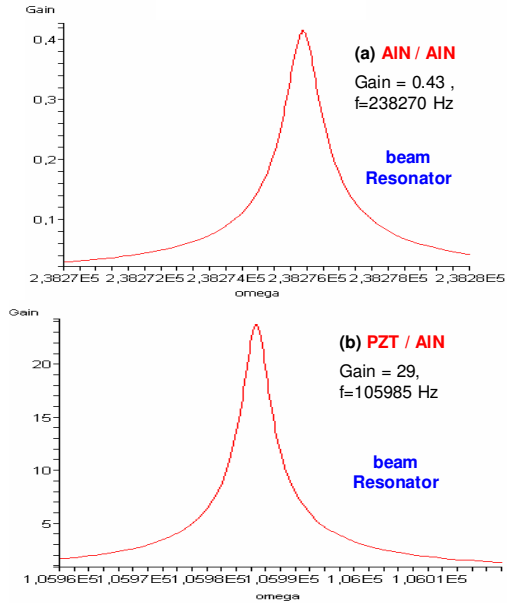

Fig. 7: Matlab simulation of bridge-type micro transformers showing the gain response around the 2nd bending mode. (a) Low gain of 0.43 is obtained when using only AlN films : (b) Higher gain of 29 is reached when using PZT at the input (primary) and AlN at the output (secondary)

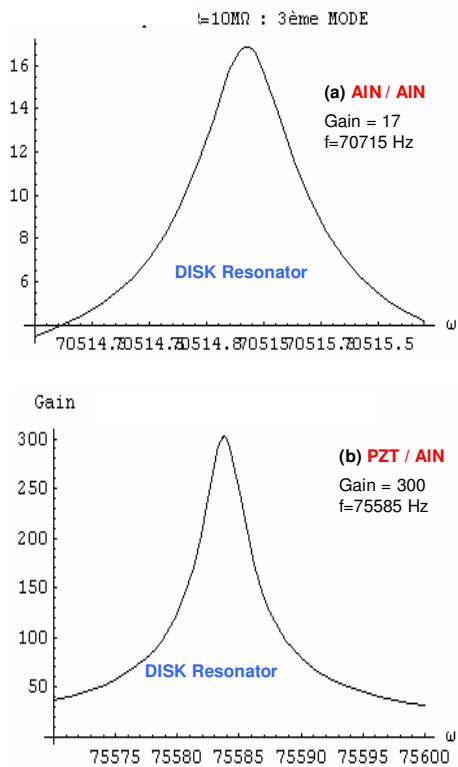

Fig 8 : Matlab simulation of disc-type micro transformers showing the gain response around the 3rd resonance mode. (a) Moderate gain of 17 is obtained when using only AlN films : (b) Higher gain of 300 is reached when using PZT at the input (primary) and AlN at the output (secondary)
Load resistance is 10 MΩ

One can note on Figs. 7 and 8 the improvement, by a factor more than 20x of the gain for the devices with two complementary in nature piezoelectric films. The deposited piezoelectric film at the vicinity of the primary electrode must have a piezoelectric coefficient $d_{31}$ much greater than the one placed on the secondary electrode. The use of different piezoelectric films at the vicinity of the primary and secondary electrodes looks promising but of course only a realization in clean room will confirm this theoretical prediction.

New geometries (shown in Fig. 9) for the micro transformer were also designed. They combine "complementary" piezoelectric layers, a geometrical shape that is aimed to increase further the strain on the secondary electrode. As these structures are open, they are also expected to decrease the effect of air damping.

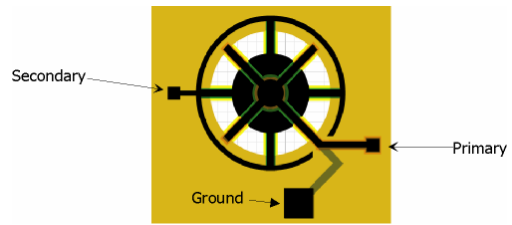

Fig.9: New resonator shapes for a Micro transformer

## V. DESIGN OF AN ELECTRONIC CIRCUIT FOR THE MICRO-TRANSFORMER

For micro transformers made with a single piezoelectric material, common to the primary and secondary and submitted to ultrasonic waves of variable frequency, the measured secondary (output) voltage V1 is in the order of 5 mV (for the resonant frequency of mode 3). However, in the case of micro transformer involving two complementary piezoelectric materials, the performed simulations predict a voltage V1 exceeding 50 mV at the secondary (output) electrode. This alternating voltage output is in any case insufficient to feed electrical power to a sensor. The use of an electronic amplification without power is therefore necessary.

The proposed electronic system (schematized in Fig. 10) was simulated with SPICE software. Its function is to amplify and generate a continuous voltage from an alternative voltage of about 50mV. This circuit consists of a SCHENKEL voltage doubler type, where diodes are replaced by N and P type MOS transistors with very low threshold voltage and also with the substrate and grid connected together.

As shown in Fig. 10, these transistors can be made on the same SOI wafer used to manufacture other electromechanical structures. They are all isolated each other by engravings of their local layers of silicon to the buried oxide layer of the SOI wafer.

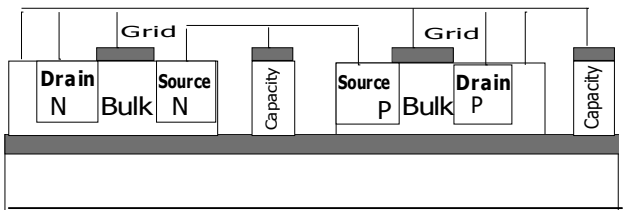

Fig 10: isolation of MOS transistors on SOI wafer

This amplifier (as shown in Figs. 11-12) operates without any power supply and allows sufficient amplification to generate up to 5.4 V DC starting from a voltage of only 50 mV AC (available at the output of the micro transformer). Also, the with power consumption is very low : 60 nW in the transient (of 2 ms) and 2.97 pW in the steady state, with the corresponding current of 60 pA at the input.

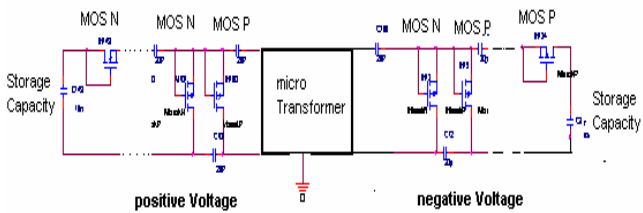

Fig. 11: The principle of electronic amplification of an AC voltage of a few tens of mV into a continuous voltage of several volts

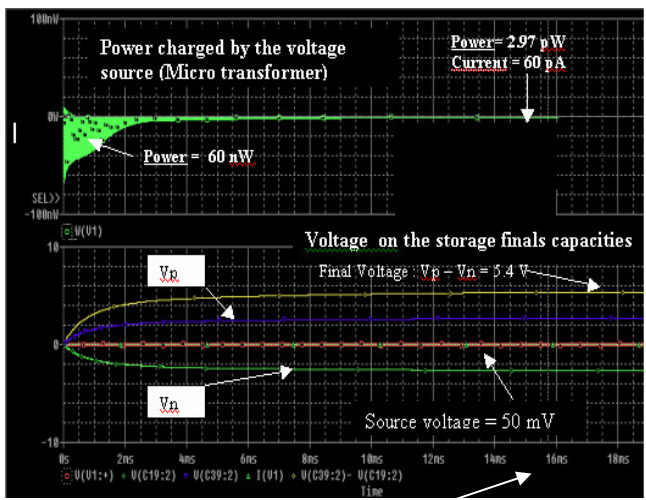

Fig 12: Spice Simulation of: 1/ power charged by the micro transformer Producing a signal (50mV, 150 kHz), number of level= 30
2/ the amplifier with a final voltage of 5.4 V, low power and current consumed by the micro transformer

## PROSPECTIVE WORK

The realization of prototypes and their characterization is still required in order to confirm the theoretical simulation studies on the second proposed version of micro-transformers involving two "complementary" piezoelectric materials. The same applies for the corresponding electrical circuit. Then we will be able to check the effectiveness of the proposed system consisting in a micro transformer, an electronic amplification circuit without power supply, a set of sensors or actuators, and a frequency modulator [5,6]. Target application of such autonomous device is based on its ability to transmit to the outside world, information from an abandoned sensor.

## ACKNOWLEDGEMENTS

The authors would like to thank Dejan Vasic, Emmanuel Sarraute and François Costa (ENS-Cachan), Eric Cattan (IEMN), Philippe Bois (CEA) for their valuable contribution to this work.

Microfabrication of the prototypes was done in the cleanroom facilities of ESIEE.

This work has the support of the French National Research Agency (ANR)in the frame of the project Ref. JC05/ 54551.

## REFERENCES

[1] D. Vasic ,E.Sarraute,F.Costa,P.Sangouard,E.Cattan , « Piezoelectric micro-transformer based on SOI structure » Sensors and Actuatirs A-physical , Volume 117 ,Issue 2,pages 317-324, 14 january 2005

[2] D. Vasic ,E.Sarraute,F.Costa,P.Sangouard,E.Cattan , "Piezoelectric micro-transformer based on PZT unimorph membrane" *Journal of Micromech. Microeng.* 14 S90-S96. September 2004.

[3] D. Vasic ,E.Sarraute,F.Costa,P.Sangouard,E.Cattan , « Piezoelectric Micro-Transformer Based on SOI structure » , MME 2003 14 st MicroMechanics Europe Worksop , 2-4 Novembre 2003 Delf, Netherlands

[4] D. Vasic , « Modélisation des micro-transformateurs » Ph.D Thesis ENS-Cachan September 2003

[5] P. Sangouard, P.Bilstein ,G.Bazin ,P .Nicole . Patents EP0P21633 ,US6137323, FR2772213, DE69814803T, JP112309025

[6] G. Bazin-Lissorgues, P. Sangouard, P. Bildstein, "Design of a Micro-mechanical capacitor with microwave functions", *EPJ Applied Physics,* N° 1, Vol. 9, pp. 75-80, January 2000.